 \documentclass[acmtog, nonacm]{acmart}

\AtBeginDocument{%
  \providecommand\BibTeX{{%
    \normalfont B\kern-0.5em{\scshape i\kern-0.25em b}\kern-0.8em\TeX}}}

\setcopyright{acmcopyright}
\copyrightyear{2023}
\acmYear{2023}

\acmConference[SC23]{Supercomputing 2023}{November 12--17, 2023}{Denver, CO}
\acmBooktitle{SC23,
 November 12--17, 2023, Denver, CO} 

\usepackage{listings}
\begin{document}

\title{Poster abstract: Quantum Task Offloading with the OpenMP API}

\author{Joseph K. L. Lee}
\email{j.lee@epcc.ed.ac.uk}
\orcid{0000-0002-1648-2740}
\author{Oliver T. Brown}
\orcid{0000-0002-5193-8635}
\email{o.brown@epcc.ed.ac.uk}
\author{Mark Bull}
\affiliation{%
  \institution{EPCC, University of Edinburgh}
  \streetaddress{47 Potterrow}
  \city{Edinburgh}
  \country{U.K.}
  \postcode{EH8 9BT}
}

\author{Martin Ruefenacht}
\affiliation{%
  \institution{Leibniz Supercomputing Centre}
  \country{Germany}}
\email{XXX}

\author{Johannes Doerfert}
\affiliation{%
  \institution{ Lawrence Livermore National Laboratory }
  \country{USA}
}

\author{Michael Klemm}
\affiliation{%
  \institution{Advanced Micro Devices \& OpenMP Architecture Review Board}
  \country{Germany}}

\author{Martin Schulz}
\affiliation{%
  \institution{Leibniz Supercomputing Centre, Technical University Munich}
  \country{Germany}
  }

\renewcommand{\shortauthors}{Lee, et al.}

\begin{abstract}
Most of the widely used quantum programming languages and libraries are not designed for the tightly coupled nature of hybrid quantum-classical algorithms, which run on quantum resources that are integrated on-premise with classical HPC infrastructure. We propose a programming model using the API provided by OpenMP to target quantum devices, which provides an easy-to-use and efficient interface for HPC applications to utilize quantum compute resources. We have implemented a variational quantum eigensolver using the programming model, which has been tested using a classical simulator. We are in the process of testing on the quantum resources hosted at the Leibniz Supercomputing Centre (LRZ).
\end{abstract}




\received{xxx}
\received[revised]{xxx}
\received[accepted]{xxx}

\maketitle

\section{Introduction}
Quantum computers come with the promise of tackling certain computational problems where the required classical computing resources scale exponentially with the problem size. This could provide advantage to several domains including chemistry, materials design, and optimization. However, they are not expected to be replacements of classical computers, but must be considered as accelerators to classical computers. Quantum computing systems work on quantum bit (qubit) states, which can be based on different technologies, including superconducting qubits, ion trap, and neutral atoms. Independent of the underlying technology, a host system is used to control the qubit states with the help of programmable control signals (typically microwaves or laser pulses) and to read and analyze the results of a quantum computation. Qubits differ from classical bits in that they carry phase information, may exist in a superposition of both computational basis states (\(|0\rangle\) and \(|1\rangle\)), and can exist in the uniquely quantum `entangled' collective states, which are not classically separable.

From the programming perspective, quantum programs consist of a sequence of gates, which are individual operations on a single or on multiple qubits. These gates are then, after optimization and translation steps, mapped to pulse sequences matching the underlying technology. In contrast to usual HPC programming approaches, where we have a clear separation in code, which is compiled ahead of time, and input data, which is added at runtime, QC systems in most cases require some form of dynamic compilation, as both code and input data impact the generated pulse sequence. Once the pulse sequences have been generated, the execution of a quantum program consists of repeated execution of these pulse sequences, each followed by measurements of the results. The result of the computation is then determined using statistical analysis to determine the most likely final state of the qubits after program execution. 

Currently the most popular way to compile and program quantum circuits is via Python libraries, such as Qiskit, Cirq, or Pennylane. These libraries then dynamically transform the gate-level description into control sequences for the quantum systems, followed by execution and measurement on the quantum system. Besides these Python libraries, there are also emerging quantum languages and intermediate representations for describing quantum circuits, including OpenQASM \cite{cross2017open} and QIR \cite{QIR}, which can facilitate optimizations and be compiled to target multiple hardware technologies and simulators. Currently most quantum resources are remote-integrated, where the quantum system is connected to a network which can be accessed via standard cloud models. A benefit of the pay-as-you-go cloud access model is that it creates a low adoption barrier.

\section{Hybrid Quantum-Classical Computing}
In the near term, given the limitations of noisy intermediate-scale quantum (NISQ) computers, hybrid quantum-classical algorithms are expected to be the main applications to run on quantum devices. 
Variational quantum algorithms (VQAs), which include Variational Quantum Eigensolvers (VQE), are a prime example of such hybrid algorithms. VQAs make use of short-depth parametrized quantum circuits, which are well suited for NISQ hardware, as well a classical variational loop. VQAs are useful for quantum chemistry simulations, optimization problems and more. However, the requirements of hybrid quantum-classical computation expose disadvantages in the above ``Python plus remote access'' model. 

\begin{figure}[h]
  \centering
  \includegraphics[width=\linewidth]{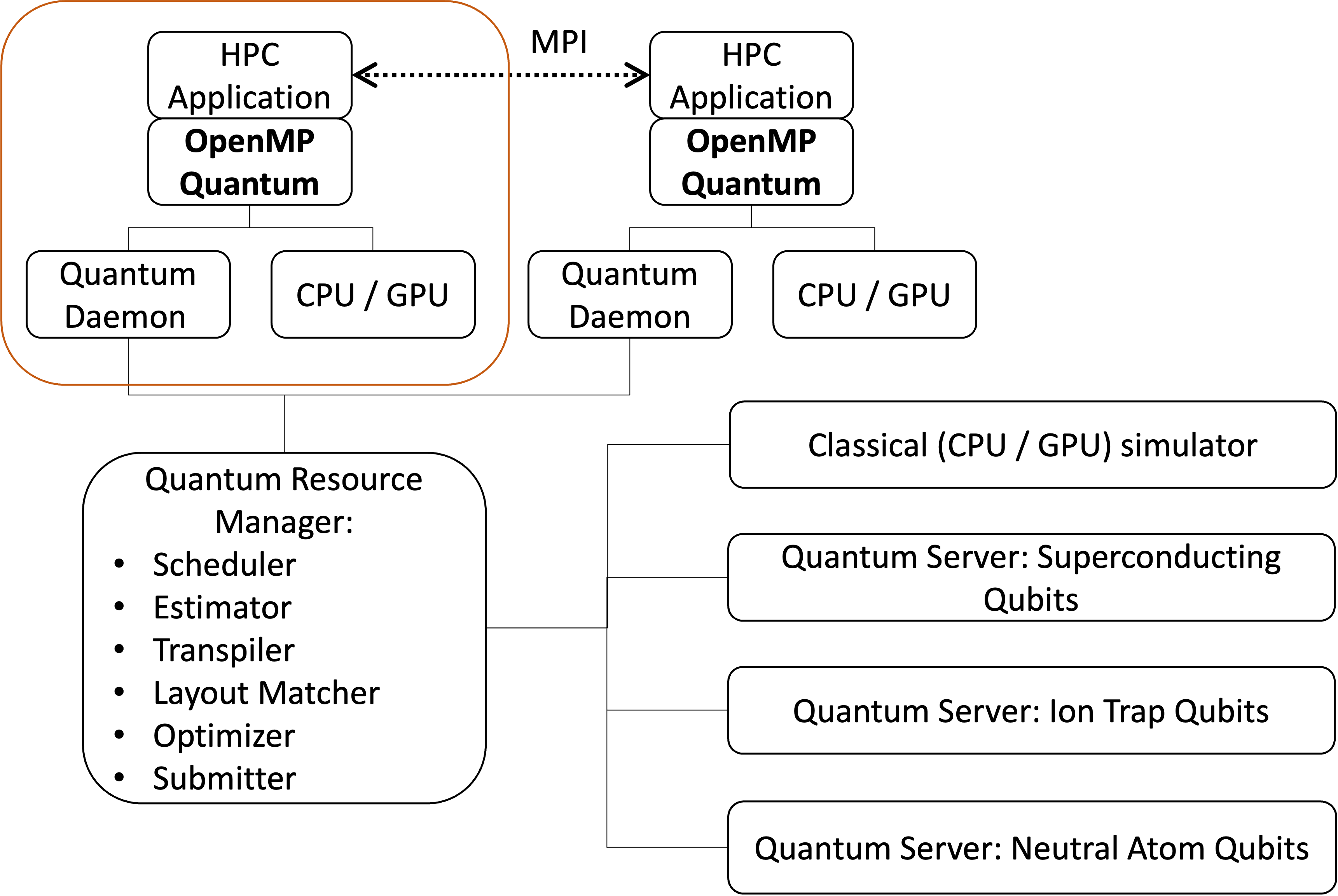}
  \caption{Architecture schematics of an HPC system with on-premise quantum resource integration.\label{fig:SystemSchematics}}
  \Description{Architecture schematics of an HPC system with on-premise quantum resource integration}
\end{figure}

Firstly, remote access incurs significant latency, which is critical for tightly coupled hybrid algorithms: the quantum and classical workloads require input from one another (e.g., parameters obtained by the classical optimizer), and message passing between the resources occurs at high frequency. Higher latencies, therefore, significantly increase the time-to-solution of these algorithms. This is why future systems, like the Euro-Q-Exa hosted by the Leibniz Supercomputing Centre, which is one of the six new EuroHPC quantum computers, will provide on-premise integration of quantum hardware, that is, where the quantum resources are located in close physical proximity to the classical compute infrastructure and are connected to the same high-speed interconnect.  Figure \ref{fig:SystemSchematics} shows a simplified architectural schematics of the integration of a quantum-classical system co-located at LRZ. 
Another drawback is that the Python interface is less favourable for the classical workload, which runs on a traditional HPC environment. The classical components of hybrid algorithms are computationally intensive, and can benefit from the optimizations provided by traditional compiled programs, usually written in C/C++ or Fortran, and parallelized with MPI and/or OpenMP. 

\section{Quantum OpenMP API}
To tackle this problem, we propose to use the OpenMP API features~\cite{OARB21} to allow offloading tasks to near-term quantum computers. The OpenMP API, which is well established in HPC programming, already provides a flexible target interface, which allows a target device to receive data from the host device, execute the control flow of a code region, and return data back to the host device. This has been widely adopted to target on-node accelerators such as GPUs; adopting OpenMP for quantum devices too can reduce potential fragmentation for applications written in C/C++ or Fortran.

The main functionality of our OpenMP extension is the ability to offload onto a quantum target. Listing \ref{lst:OmpTargetC++} shows a simple example of a C program to repeatedly construct and measure one of the Bell states within a quantum target region. Bell states are maximally entangled two-qubit states~\cite{Bell_1964}, and are often used to determine that a device or experiment is really capable of preparing quantum states.

\begin{lstlisting}[numbers=left,language=C++,caption={Bell state creation and measurement with OpenMP},label=lst:OmpTargetC++]
void bell_0() {
    int states = 4; 
    int shots = 1000;
    int results[states];
    #pragma omp target map(from:results)
    {
        omp_q_reg q_regs = omp_create_q_reg(2);
        omp_q_h(q_regs, 0);
        omp_q_cx(q_regs, 0, 1);
        omp_q_measure(q_regs, shots, result);
    }   
}
\end{lstlisting}
By default, the target regions are synchronising, so the OpenMP thread will block until the device code has completed execution and any associated data returned to the host. This model is suitable for tightly coupled hybrid algorithms. On the other hand, asynchronous execution can be enabled by adding a \texttt{nowait} clause to the target directive. This can be useful if the quantum device is busy and classical computation can still be performed.

The proposed implementation contains function calls to create and measure quantum registers, and apply a standard set of single and two qubit gates. This will then be transpiled into QASM or QIR (Listings \ref{lst:QASM} and \ref{lst:QIR}), which can subsequently be passed onto the quantum resource manager for further optimization and scheduling onto the required quantum device or simulator. Alternatively, the user will also be able to directly program using QASM.

\begin{lstlisting}[numbers=left,escapeinside={(*}{*)},language=C,caption={QASM for Bell state},label=lst:QASM]
OPENQASM 2.0;
include "qelib1.inc";
qreg q[2];
creg c[2];
h q[0];
cx q[0],q[1];
measure q (*$\to$*) c;
\end{lstlisting}

\begin{lstlisting}[numbers=left,language=C,caption={QIR for Bell state},label=lst:QIR,breaklines=true]
define void @main() #0 {
entry:
  call void @__quantum__qis__h__body(%Qubit* null)
  call void @__quantum__qis__cnot__body(%Qubit* null, %Qubit* inttoptr (i64 1 to %Qubit*))
  call void @__quantum__qis__mz__body(%Qubit* null, %Result* writeonly null)
  ...
  ret void
}
\end{lstlisting}

We have implemented a VQE algorithm using OpenMP target offload, which has been tested using a classical simulator. We are currently in the process of testing on the physical superconducting
qubits hosted by LRZ.

\section{Acknowledgements}
Part of this research has been conducted as part of the Munich Quantum Valley (MQV), which is supported by the Bavarian state government with funds from the Hightech Agenda Bayern Plus. Further funding comes from the German Federal Ministry of Education and Research (BMBF) through the project MuniQC-SC/ATOMS as well as Q-Exa.

The OpenMP name and the OpenMP logo are registered trademarks of the OpenMP Architecture Review Board. Other names and brands may be claimed as the property of others.
\bibliographystyle{ACM-Reference-Format}
\bibliography{QOMP}


\end{document}